\documentclass{article}
\pdfoutput=1
\usepackage[utf8]{inputenc}
\usepackage[english]{babel}
\usepackage[T1]{fontenc}
\usepackage{amsmath}
\usepackage{hyperref}
\usepackage{arxiv}
\usepackage{amssymb}
\usepackage{tikz}
\usepackage{lipsum}
\usepackage{mathtools}
\usepackage{physics}
\newtheorem{theorem}{Theorem}

\begin{document}
\title{{An Extended Quantum Process Algebra (eQPAlg) Approach For Distributed Quantum Systems}} 

\author{
  Salman Haider\\
  Department of Computer Science\\
  Government College University\\
  Anarkali Bazaar 54000 Lahore, Punjab, Pakistan \\
  \texttt{ravian.salman@hotmail.com} \\
   \And
 Dr. Syed Asad Raza Kazmi \\
  Department of Computer Science\\
  Government College University\\
  Anarkali Bazaar 54000 Lahore, Punjab, Pakistan \\
  \texttt{arkazmi@gcu.edu.pk}  \\
  }


\maketitle
\begin{abstract}

In this work, we have expounded the communication procedure of quantum systems by means of process algebra.  The main objective of our research effort is to formally represent the communication between distributed quantum systems.  In this new proposed communication model we have ameliorated the existing rules of Lalire's quantum process algebra QPAlg.  We have brought some important modification in QPAlg by introducing the concept of formally specifying the Quantum teleportation protocol.  We have further introduced the formal description of protocol by using programs that best explains its working and satisfies the specification.  Examples have been provided to describe the working of the improved algebra that formally explain the sending and receiving of both classical as well as quantum data, keeping in mind the principal features of quantum mechanics.
\\
\\
{\normalfont\textbf{Keywords}: Quantum computing, Quantum mechanics, Quantum teleportation protocol, Process algebra}
\end{abstract}

\textbf{\section{Introduction}}

Quantum information theory is an advanced approach and is basically an amalgam of computer science, physics, and mathematics. It uses the principles of quantum mechanics. In quantum mechanics, the state of the system is expressed by a wave function. The fundamental laws of quantum mechanics are widely adopted in computation as well as in communication. As a comparatively modern computational model, the quantum computing\cite{42}\cite{43} brings the dawn of solving the so-called NP problem because of the strong parallel computation power of quantum computing. Many of the principles of quantum mechanics, such as quantum no-cloning theorem, uncertainty principle, and entanglement provide quantum communication protocols the provable security. We first introduce the key aspects of quantum mechanics in computer science such as quantum computation and quantum communication. The novel concepts of process algebra in the light of formal methods have been explained further with a brief overview of previous researches.

\textbf{\subsection{Quantum Computation \& Communication}}
Quantum Computing is a new computational theory in computer science. The machine which carries out this computation is referred to as Quantum Computer. We know in digital computing, computations are performed by Standard Computers that consist of transistors. Such types of standard computers are able to interpret data only in the binary format that consist of either 0 or 1 state. On the other hand, the novel Quantum computation uses quantum bits (qubits) which can be in a superposition of the two states. Quantum computers could possess incredibly large processing power and could process much more amount of information at a given time. The amount of information flow quantum mechanical computer can measure would be: $2^{N}$ where N = Number of qubits. For larger qubits, if we are given 300 qubits then $2^{300}$ would be huge enough that it might be equal to the total number of particles present in the observeable universe. \\

A classical computer can take millions of years to find the prime factor of 2048 bits number whereas using a quantum computer it will take only a few minutes. The processing power of quantum computer when we are given with 30 qubits would be equal to the classical computer that computes at 10 Teraflops. These computers took advantages of quantum superposition to reduce \textbf{T(n): Number of steps requires to get the computational results} and can only perform faster operations only if we are using some particular algorithms as we can perform a number of computational operations in an exponentially smaller time.
Nowadays, communication devices and protocols are based on the principles of classical physics where information is communicated in the form of bits. In addition to quantum computing, the important aspects of quantum mechanics \cite{12}  such as quantum teleportation\cite{5}, no-cloning theorem and entanglement \cite{13} needs to be studied (For details, one should refer to these \cite{12}\cite{13}\cite{42}\cite{43}. Quantum computing actually offers the option of an entirely new paradigm for information processing, with the alluring possibility of having the capacity to break existing crypto-systems.\\

In recent years substantial progress has been made in the area of quantum communication. Moreover, the laws of quantum mechanics have made it possible to accomplish the communication tasks that are not possible using classical principles. Some of these tasks include dense coding \cite{14}; unconditionally secure quantum key distribution \cite{12} and quantum teleportation (QT). The basic protocols of quantum communication tasks were propounded from 1984 to 1993. For example
\begin{itemize}
    \item In 1984, Bennett and Brassard put forward the very first protocol of QKD known as BB84 protocol. In this protocol, there is a sender that relays a random sequence of bits and then distributes it to a remote receiver using quantum means.
    \item Moving to the concept of dense coding, aka super dense coding was proposed first time by Bennett \cite{3} in 1992. In this, a sender can communicate two classical bits of information to the receiver by sending only one qubit of information with the condition that the sender and receiver must share a prior entanglement.
    \item In 1993, Bennett introduced the quantum teleportation (QT) scheme \cite{5}\cite{13}\cite{15} in which a sender transmits an unknown quantum state to a distant receiver using two bits of classical communication along with an entangled state that has already been shared by both the sender and receiver.
\end{itemize}
After the publication of these fundamental research works regarding quantum protocols, several new quantum communication protocols have been coined. Some of which do not need security as these are already secured such as quantum teleportation and super dense coding and some of these require security proof such as protocols of quantum key distribution (QKD), Deterministic secure quantum communication (DSQC), Quantum secret sharing (QSS), quantum dialogue (QD) and Quantum secure direct communication (QSDC)  \cite{16}\cite{17}.

\textbf{\subsection{Algebra of Quantum Processes}}
In the algebra of quantum processes, formal methods are used to provide systematic techniques for modeling  \cite{5}, analysis and verification of quantum systems. It is normally used to model the behavior of systems that are created with both classical and quantum information. There are many formal languages available pertinent to specification and verification. There has been greater than before interest in process algebra \cite{11} for the specification and modeling of concurrent systems. It is appealing to unite quantum computing and classical computing under the same process algebra framework as most quantum communication protocols involve quantum information and classical information, quantum computing and classical computing. In the last few years, serious efforts have been made to link the computer science areas like programming languages and formal verification with quantum information processing.
\\
The scheme of devising a robust computational machine based on quantum mechanical laws dates back to 1980 \cite{39} and \cite{45}. Then, Feynman \cite{19} propounded that a quantum mechanical system becomes useful for computing, a great deal of attention was paid over quantum computing based on quantum information theory. Quantum computing (as discussed earlier) can provide a great speedup as compared to its classical analogue after taking advantage of the characteristics of being in a superposition of quantum states. \cite{25} \cite{26} \cite{27}.
In order to offer an approach of examining the computational difficulties in a more logical way, some researchers started studying the semantics and design of quantum programming languages (QPL). Knill composed a set of fundamental transition rules for describing the quantum pseudo-codes \cite{14}, whereas Omer proposed the very 1st real-time programming language for quantum systems known as QCL\cite{25} \cite{26}. In the meantime, different programming languages have been proposed that would possibly be working on quantum computers (Refer to this survey by Gay \cite{46}), among which is qGCL which resembles Dijkstra guarded commands language was formulated by Zuliani et al. in \cite{28} \cite{29}. An extended version of C++ with principles of quantum systems, which is being proposed as a C++ library, was proposed by Bettelli \cite{24}. But the first breakthrough was when Selinger \cite{37} proposed the quantum language (QPL) which was first function-based programming language. QPL combines high-level classical design with its basic operations on quantum information. A short time ago, Selinger et al. \cite{41} further enhanced this function-based programming language with the same quantum information paradigm along with classical control where the information is being shared, the same idea has been extended in the calculus being introduced by Arrighi et al. \cite{40}.  \\
\\
Van \cite{38}\cite{39} presented a quantum $\lambda$ calculus collaborating higher-order logic. Up till now the languages in hand, mostly serve the sequential quantum computing, where no transition rules have been presented to model the communication between distant observers. The formulation of such programming languages has now started that would possibly describe the concurrent quantum systems along with the principles of communication between them.
\\
Quantum cryptography \cite{9}, is able to provide an utmost level of security even if it is vulnerable to attacks of intended eave-droppers. Rapid progress was made in quantum cryptographic systems that now these have become commercially available \cite{36}. Now we are in direct need of a language which can describe a concurrent system more instantly both on the basis of quantum and classical computation. Additionally, a system for formulating the quantum concurrent activities is needed that can provide methods to verify the various properties of that specific system such as security, secrecy, and correctness of any cryptography-based quantum protocol. Jorrand et al. \cite{32}, and Gay et al. \cite{31} independently made the effort of designing a generalized framework of modeling concurrent quantum systems where process algebra has been extended to the quantum setting. Jorrand et al. \cite{32} formulated an algebra for quantum systems which could describe both quantum and classical data sharing. After that, Lalire proposed a probabilistic branching bisimulation for their previously formulated language which was able to identify quantum systems connected to their process graph with similar branching structure\cite{34}. While Gay et al. \cite{4} described a process algebraic language called Communicating Quantum Processes (CQP) which is an instance of calculus that was introduced by Lalire et al. \cite{34}, which gave both the primitives for unitary transformations and measurements for the analysis of quantum communication systems.
\textit{\subsection{Overview of this paper}}
This paper is organized as follows: In \textbf{Section 2}, we review some basic notions from linear algebra and quantum mechanics which will be used in this paper. The syntax of eQPAlg is presented in \textbf{Section 3}. \textbf{Section 4 and 5} are the major contribution of the present paper. First, we write the formal specifications of the well-known quantum teleportation protocol with eQPAlg then show that it indeed teleports any qubit from one party to another with a cost of sending only 2 classical bits. Secondly, programs are written that could satisfy the specification and these programs can also be implemented in the form of computer code for simulation. The improved algebraic transition rules for communication between two quantum processes are defined and examples are given for supporting these transition rules. \textbf{Section 6} is the concluding section in which we outline the research work and point out some problems for further study.
\textbf{\section{Preliminaries}}
For reader's understanding, we introduce some fundamental concepts about linear algebra and quantum mechanics (Reader is intended to read \cite{42}\cite{43} and \cite{44} for details). The syntax and structural operational semantics of the former quantum process algebra can be found here (\cite{3}\cite{32} and \cite{34} for detailed study).

\textbf{\subsection{Linear Algebra: Vector Spaces and Operators}}
Study of vector spaces and linear transformations in linear algebra. Hilbert Space, where quantum mechanics are formulated, is an significant vector space in mathematics and physics. First we give fundamental definitions and notes, then officially define Hilbert Space.

\textbf{\subsubsection{Qubit}}
The fundamental unit of representing information in quantum computation is called \textit{quantum bit} or \textit{qubit}\cite{1}. Alike \textit{bit}, a qubit can be in one of the two states. We will write these states by $\ket{0}$ and $\ket{1}$. Anything enclosed using this notation $\ket{\,}$ is known as \textit{state}, \textit{vector} or a \textit{ket}. A classical bit can only be in one state, it can be 0 or 1. A qubit can exist in $\ket{0}$ or $\ket{1}$ and it can also occurs in both states and this state is called \textit{superposition}. If we have a state $\ket{\psi}$ then the superposition state will be:
\\
\begin{equation*}
    \ket{\psi} = \alpha\ket{0} + \beta\ket{1}
\end{equation*}
$\alpha$ and $\beta$ are the complex coefficients. While measuring $\ket{\psi}$, we can have the probability of finding $\ket{\psi}$ in one of the two states which is calculated by modulus squared $\alpha$, $\beta$ such that:
\\
\\
$|\alpha|^2$: Gives the probability of finding $\ket{\psi}$ in $\ket{0}$ 
\\
$|\beta|^2$: Gives the probability of finding $\ket{\psi}$ in $\ket{1}$ 

\textbf{\subsubsection{Vector Spaces}}
The term vector is used for representing the quantum state of a physical system in a \textit{complex} vector space. An important vector space in quantum computation is $\mathbb{C}^n$ consisting of "n-tuples" of different complex numbers. We can name the components of $\mathbb{C}^n$ using $\ket{a}\, \ket{b}\, \ket{c}$. One can denote each part of this type of vector space in the form of n-dimensional \textit{column vector} such as:
\begin{align*}
    \ket{a} = 
\begin{pmatrix}
  a_1\\
  a_2\\
   .\\
   .\\
   .\\
  a_n\\
\end{pmatrix}
\end{align*}
We use such notations to denote qubits. A qubit in Hilbert space $\mathbb{C}^2$ (which we will discuss later) can be written in column vector format:
\begin{align*}
\ket{\psi} =
\begin{pmatrix}
    \alpha \\
    \beta  \\
\end{pmatrix}
\end{align*}

\textbf{\subsubsection{Basis and Dimension}}
If we have a group of vectors spanning the space of a vector V and linearly independent to each other, we can call this group as a \textit{basis} and the No. of elements in this basis set is known as the \textit{dimension} of V \cite{1}. Basis set of qubits states $\ket{0}$ and $\ket{1}$ for $\mathbb{C}^2$ with:
\begin{align*}
    \ket{0} = 
    \begin{pmatrix}
        1\\
        0\\
    \end{pmatrix}
    ,\; \ket{1} = 
    \begin{pmatrix}
        0\\
        1\\
    \end{pmatrix}
\end{align*}

\textbf{\subsubsection{Inner Product}}
The generalized form of dot products that can be along with simple vectors in the Euclidean space. In dot product we have two vectors mapped into a real number whereas inner product get two vectors from $\mathbb{C}^2$ and do the mapping into complex number so the inner product is normally considered as complex number. We have two vectors $\ket{u}$, $\ket{v}$ and the inner product can be represented by $\braket{u}{v}$. To calculate inner product among two vectors, we have to first compute the hermitian conjugate of one of the two vectors.
\begin{align*}
    (\ket{u})^\dag = \bra{u}
\end{align*}
If we have
\begin{align*}
    \braket{u\,}{\,v} = 0 
\end{align*}
then it is stated that $\ket{u}$, $\ket{v}$ will be \textit{orthogonal} to each other and if the norm of a vectors is unity such that
\begin{align*}
    \braket{u}{v} = 1
\end{align*}
then $\ket{u}$, $\ket{v}$ are \textit{normalized}. If we have a vector which isn't normalized then we get a normalized vector by calculating the \textit{norm} (that can be a real number) this way, 
\\
\begin{equation*}
    ||u|| = \sqrt{\braket{u}{u}}
\end{equation*}
and dividing the vector by it. If every component of a group of vectors is orthogonal as well as normalized to one another then this set is said to be the \textit{orthonormal}. 

\textbf{\subsubsection{Bra-ket Notation}}
Most commonly written notation for representing the states of a quantum system in the orthonormal basis is \textit{Bra-ket}. If we have vectors $\ket{u}$, $\ket{v}$ then we will call it a \textit{ket} or \textit{column vector} and we can find its \textit{bra} or \textit{dual vector} by calculating the \textit{hermitian conjugate}. So, the bras corresponding to kets will be $\bra{u}$, $\bra{v}$.

\textbf{\subsubsection{The Trace of an operator}}
If we have an operator in the form of matrix then we can compute the \textit{trace} of this operator by summing up its diagonal elements. Such as,
\begin{align*}
Z = 
  \begin{pmatrix}
    w & x \\
    y & z \\
\end{pmatrix}, \quad  Tr(Z) = w + z
\end{align*}
In case, the operator can be in outer product form then we can calculate the trace by adding the inner products with its the basis states. If we have our vector in basis $\ket{v_i}$, then trace of operator will be\\
\begin{equation*}
    Tr(Z) = \sum_{i=0}^n \expval{v_i}{Z}
\end{equation*}
\textbf{\subsubsection{Hilbert Space}}
An Hilbert Space\cite{1} is a two dimensional complex vector space and a generalization of \textit{Euclidean space} where a qubit resides. It is represented by $\mathcal{H}$. 
\textbf{\subsubsection{Tensor Product}}
If $H_1$, $H_2$ are 2-Hilbert spaces where $\ket{\phi_1} \, \epsilon \, H_1$  $\ket{\phi_2} \, \epsilon \, H_2$ are the two vectors belongs to $H_1$ and $H_2$ then the \textit{tensor product} for these vectors can construct a larger Hilbert space H represented by state vector $\ket{\psi}$ such as
\begin{equation*}
    \ket{\psi} = \ket{\phi_1 \otimes \phi_2}
\end{equation*}
\textbf{\subsection{Quantum Mechanics}}
Now we provide a review on the basics of quantum mechanics that is needed to understand the paper.
\textbf{\subsubsection{Quantum Gates}}
In the context of quantum mechanics, data processing id done through gates also known as unitary operators\cite{1}. A quantum gate or a logic gate is the basic element of quantum computation consisting of small quantum registers and connected with wires to build a circuit. Quantum algorithm is actually the number of gates which are connected together and performing different operations on quantum states to get the desired measurement. They can also be represented by \textit{unitary matrices}. Quantum gates can perform reversible computation whereas in most of the classical gates only perform irreversible computation. For examples, a classical \textit{AND} gate performs operation on two different bits and the measurement results will not be reverted back to get the original bits. However, in reversible computing such as \textit{NOT} gate, one can measure the input bit from the resulting bit.    
\subsubsubsection{Hadamard Gate:}
Hadamard gate\cite{1} can perform actions on a single qubit. This plot the basic state $\ket{0}$ to $\frac{\ket{0} + \ket{1}}{\sqrt{2}}$ and $\ket{1}$ to $\frac{\ket{0} - \ket{1}}{\sqrt{2}}$ that states that the measurement to be performed will have equal probabilities of 0 or 1. It is represented as the 2 X 2 matrices form\cite{1}:
\begin{align*}
    H = \frac{1}{\sqrt{2}}
    \begin{pmatrix}
        1 & 1 \\
        1 & -1
    \end{pmatrix}
\end{align*}
As, $HH^{*} = I$, here I is the identity matrix and hence H is a unitary operator.

\subsubsubsection{CNOT Gate:} 
CNOT gate operates on a pair of qubits where the first qubit is the control qubit whereas the other qubit is the target qubit\cite{1}. NOT Gate or Pauli-X operation will be applied to target qubit with respect to control qubit values. If the value of control qubit is $\ket{1}$ then it flips the state of target qubit to $\ket{0}$ otherwise leaves it unchanged. The matrix representation of a 2-qubit CNOT operator can be given by
\begin{align*}
    CNOT = 
    \begin{pmatrix}
        1 & 0 & 0 & 0 \\
        0 & 1 & 0 & 0 \\
        0 & 0 & 0 & 1 \\
        0 & 0 & 1 & 0 \\
    \end{pmatrix}
\end{align*}

\subsubsubsection{Phase-Shift Gates}
This a group of single-qubit gates that changes the basic state $\ket{1}$ to $e^{\iota \phi}\ket{1}$ and leaves the basic state $\ket{0}$ unchanged. The phase-shift is represented by $\phi$ 
\\
\begin{align*}
    R_\phi = 
    \begin{pmatrix}
        1 & 0 \\
        0 & e^{\iota \phi} \\
    \end{pmatrix}
\end{align*}
The examples of phase-shift gates are Pauli-Z gate where $\phi = \pi$, the phase gate where $\phi = \frac{\pi}{2}$ and $\frac{\pi}{8}$ gate where $\phi = \frac{\pi}{4}$ 
\textbf{\subsubsection{Pauli Operators}}
Pauli operators are of fundamental importance in quantum computation. Different symbolic representations are used for Pauli operators, sometimes represented by $\sigma_0  \, , \sigma_1 \, , \sigma_2 \,$ , $\sigma_3 \, $ and sometimes I, X, Y or Z, or $\sigma_0 \, , \sigma_x \, , \sigma_y \,$ , $\sigma_z \, $ where, $\sigma_0 = I  \, , \sigma_1=\sigma_x = X \, , \sigma_2 =\sigma_y = Y \,$ and $\sigma_3 = \sigma_z = Z \, $ \cite{1}. As the name suggests, applying identity on any state will leaves the state unchanged: 
\begin{equation*}
    I\ket{\psi} = \ket{\psi}\quad \text{such that} \quad I\ket{0} = \ket{0}, \:  I\ket{1} = \ket{1}
    \label{eq:3.4}
\end{equation*}
The next operator is known as \textit{bit-flip} and is denoted by X. It operates the same as \textit{NOT operator}:
\begin{equation*}
    X\ket{0} = \ket{1}, \; \;  X\ket{1} = \ket{0}
    \label{eq:3.5}
\end{equation*}
Third operator is abbreviated by Y and it operates as follows:
\begin{equation*}
    Y\ket{0} = -\iota\ket{1}, \; \;  Y\ket{1} = \iota\ket{0}
    \label{eq:3.6}
\end{equation*}
Lastly, we have Z operator sometimes called \textit{phase-flip} and is denoted by Z. It acts as follows:
\begin{equation*}
    Z\ket{0} = \ket{0}, \; \;  Z\ket{1} = -\ket{1}
    \label{eq:3.7}
\end{equation*}
\textbf{\subsubsection{Hermitian Operator}}
In quantum theory, one special type of operator is \textit{Hermitian}. Operators which represents the physical observables are hermitian. An operator A is said to be hermitian if and only if
\begin{equation*}
    A = A^\dag
\end{equation*}
The Pauli operators are hermitian. 
\textbf{\subsubsection{Unitary Operator}}
A linear operator represented by \textit{U} also known as \textit{unitary}, if the adjoint of \textit{U} equals its inverse if $UU^{\dag}\, = \, U^{\dag}U \, = \, I$, here I represents the identity operator and $U^{\dag}$ denotes the conjugate transpose\cite{1}.
\textbf{\subsubsection{Density Operators}}
For describing quantum state to vector we can use use Density operators. Density operators represents the mixed(uncertain) state of the system in a compact way. There are basically two formulations of density operators, first one formulation uses \textit{probabilities} and other uses \textit{trace-preserving} matrices. A density operator is a linear operator in Hilbert space $\mathcal{H}$ and is represented by $\varrho$. Density operator satisfies following conditions:\\
\begin{enumerate}
    \item A density operator is Hermitian, that is, $\varrho$ = $\varrho^\dag$.
    \item $\varrho$ must be positive, only if, $\expval{\psi}{\varrho}\geq 0$.
    \item Trace of density operator $\varrho$ is equal to one, that is, $Tr(\varrho) = 1$.
\end{enumerate}
$\mathcal{D}(\mathcal{H})$ represents the group of all positive density operators exists in $\mathcal{H}$.

\textbf{\subsubsection{Quantum Measurement}}
Quantum measurement contains of a group of measurement operators which is $M_{m}$, m represents the measurement results and satisfying the following: 

   $$ \sum_{m} M^\dag_{m} M_{m} = \mathcal{I}_\mathcal{H} $$

\textbf{\section{Algebra for Distributed Quantum Systems - eQPAlg}}

Motivated from classical process algebras, which offer a framework for modeling cooperating computations, a process algebraic notation has been outlined, named QPAlg for Quantum Process Algebra, that provides a uniform style to formal descriptions of synchronous and distributed computations comprising each quantum and classical elements.

On the quantum aspect, QPAlg provides quantum variables, operations on quantum variables (unitary operators and measure observables), likewise as new sorts of communications involving the quantum world \cite{3}.
\\
The idea of this paper is to model the quantum operations involved in distributed quantum systems into the algebra of Lalire \cite{16} with a slight extension of writing specifications. The extended algebra also provides the support not only to formally analyze the data flow in quantum communication protocols that involve classical data, but also we can write down their specifications. The current work introduces an easy to understand and a short description of classical and quantum components as well as communication between distributed quantum systems.

\textbf{\subsection{The Quantum Process Algebra - eQPAlg}}
\texttt{Syntax:} 
\begin{equation*}
\begin{split}
 process ::= nil \,|\, end \,|\, action.process\,|\,  process;process \,|\, process\backslash\{gate\_list\}    \\
    |\, \textbf{[} var\_decl\_list.process \textbf{]}   \,|\, process \textbf{[} var\_list \textbf{]}
 \,|\,process \parallel process \,|\,process\parallel_{\{var\}}process
\end{split}
\end{equation*}

\begin{equation*}
\begin{split}
 action \,\, ::= \,\, com\,|\, unit\_transf  |\, measure
\end{split}
\end{equation*}

\begin{equation*}
\begin{split}
 com \,\, ::= \,\,  gate ! exp \,|\, gate ! measure  |\, gate ? variable
\end{split}
\end{equation*}

\begin{equation*}
\begin{split}
 unit\_transf \,\, ::= \,\,  unitary\_operator \textbf{[} var\_list \textbf{]}
\end{split}
\end{equation*}

\begin{equation*}
\begin{split}
 measure \,\, ::= \,\,  observable \textbf{[} var\_list \textbf{]}
\end{split}
\end{equation*}

\begin{equation*}
\begin{split}
 var\_decl \,\, ::= \,\,  variable : var\_type 
\end{split}
\end{equation*}

\begin{equation*}
\begin{split}
 proc\_def \,\, ::= \,\, process\_name \, \xrightarrow{def} \, process
\end{split}
\end{equation*}

\begin{equation*}
\begin{split}
 Spec \,\, ::= \,\, \big(Var,Op,Eq,State\big)
 \end{split}
\end{equation*}

Where: \\
\begin{equation*}
Op \,\, ::= \,\, \wedge \,|\, \vee \,|\, \Rightarrow \,|\, + \,|\, \geqslant \,|\, \coloneqq \,|\, \equiv
\end{equation*}
    Var ::= Any variable of classical or quantum type \\
    Eq ::= Any equation or expression \\
    State ::= Any quantum state represented by Dirac Notation
\newline
\newline
\begin{theorem}[Specification] 
A specification is a statement written using mathematical expressions that is used to describe a system during the system analysis, system design and requirement analysis. A specification is generally not an executable program. A specification is said to be implementable if there is at least one output state produced for each input state of the specification. They are used to describe the what, not the how. Normally, in writing specifications various process calculi are used. However, we will be using standard logical expressions such as conjunction $\wedge$ for joining expressions, disjunction $\vee$ in place of OR, implication $\Rightarrow$ for implication, equivalence $\equiv$ for approximately equal to. 
\end{theorem}
\begin{theorem}[Program]
To specify the  behavior of a computer a program is used or A program is an implementable form of specification. 
\end{theorem}
\textbf{\subsection{Representing Distributed Quantum Systems}}
Let we have two quantum processes $P$ and $Q$ that represent two quantum systems in the state $\ket{\psi}$ and $\ket{\phi}$ then $P \parallel Q$ represents the distributed quantum systems that may have entanglement between them then we have: 
\begin{itemize}
    \item p and q are integer variables representing the variables associated to $P$ and $Q$. 
    \item $Var_{p}$ and $Var_{q}$ are the set of variables used by $P$ and $Q$.
    \item If we have $n+m$ quantum system in state $\psi$ then first n\-qubits represents quantum system $P$ and m\-qubits represents quantum system $Q$ then 
    \begin{equation*}
        P = \psi_{0...n} \quad Q = \psi_{n...m}
    \end{equation*}
\end{itemize} 
If we have
\begin{equation*}
   \psi = \frac{\ket{00}+\ket{11}}{\sqrt{2}}
\end{equation*}
as an entangled quantum system then $P$ owns the first qubit of state $\psi$ and $Q$ owns the second qubit of state $\psi$ and we can say that the entangled system $\psi$ has been distributed as well as shared between  $P$ and $Q$. 
\newline\newline
To represent such entangled quantum systems, we use parallelism:
$$ P\,\parallel_{\psi}\, Q = (P \otimes Q)_{\psi}$$
where 
$$ P\,\parallel_{\psi}\, Q = \psi_{0}, p \otimes \psi_{1},q $$
So, $$ P\,\parallel_{\psi}\, Q= Var_{p} \otimes Var_{q} $$
\begin{theorem}Distributed Quantum Systems are both Distributed and Shared \\
\textbf{Shared:}
If $P$ performs any operation to its state $\psi_{0}$ then it will affect the results of $Q$'s $\psi_{1}$ state.\\
\textbf{Distributed:}
$P$ and $Q$ can only perform the operation to its own qubit and can access their own set of variables. 
\end{theorem}
\textbf{\subsubsection{Performing Operations on Quantum Systems}}

The representation of unitary operations is written as:
     \begin{equation*}
        P = \psi_{0...n} \quad Q = \psi_{n...m}
    \end{equation*}
Then 
 \begin{equation*}
        P = \cup_{{P}_{\psi_{0...n}}} \quad Q =\cup_{{Q}_{\psi_{n...m}}}
    \end{equation*}
For Distributed Quantum Systems: \newline
To apply a unitary operation on two distributed quantum systems, we write it as follows:
\begin{equation*}
 \begin{split}
 P\,\parallel_{\psi}\, Q = (\cup_{{P}_{\psi_{0...n}}} \otimes \cup_{{Q}_{\psi_{n...m}}}) \\ 
 = (\cup_{{P}} \otimes \cup_{{Q}})_{\psi} \quad \quad \quad \; \\
 \text{where} \quad \psi = (\psi_{0...n} \otimes \psi_{n...m}) \quad \quad
\end{split}
\end{equation*} 
\textbf{\subsubsection{Performing Measurement on Quantum Systems}}

The representation of Quantum measurement is written as: 
\begin{equation*}
 \begin{split}
P = measure\;P_{\psi,p} \quad  Q =  measure\;Q_{\psi,q} 
\end{split}
\end{equation*} 
For Distributed Quantum Systems: \newline
To apply measurement on two distributed quantum systems, we write it as follows:
\begin{equation*}
 \begin{split}
P\,\parallel_{\psi}\, Q =  measure\;{P}_{\psi_{0},p} \otimes measure\; Q_{\psi_{1},q} 
\\
\psi_{0} =  \psi_{0...n} \, \text{qubit system,} 
\\
\psi_{1}=\psi_{n...m} \, \text{qubit system}
 \end{split}
\end{equation*}
\textbf{\section{Formal Analysis of Quantum Teleportation}}
We have used Quantum teleportation which is a quantum communication protocol initially described by Bennett et. al. (\cite{15}), Now this protocol has become so famous and is widely used as an integral part of some sophisticated quantum communication protocols.We have benefited from this advanced protocol because it helps to achieve successful passing of quantum data using a classical channel along with a pair of entangled qubits, which implies that no qubits could be transferred during this process. 
\\
Informal description of Quantum teleportation protocol\cite{15} is as follows: In a state $\ket\phi$ = $\frac{\ket{00} + \ket{11}}{\sqrt{2}}$ , Alice and Bob have a common pair of entangled qubits where Alice having some qubits with some unidentified state that she wants to establish communication with Bob's quantum state $$\ket\psi = (\alpha\ket{0} + \beta\ket{1})$$. Alice initiates it by associating the qubit with the first half of the entangled pair that she actually intends to teleport. To do so, Alice first has to perfom CNOT, then Hadamard Transformation and apply measurement operator on the qubits she has. Alice then shares the results of this measurements to Bob. Here the measurement result will be the 2-bits. Bob then gets these 2\-bits and starts performing identity operator to his qubit. Bob ultimately retrieves the state that Alice wants to teleport to him. 
\\
The informal description of working of quantum teleportation  can be seen in the Figure 4.1. The description that has been presented above is not enough because it just defines how the quantum system evolves. 
\begin{figure}[!htb]
    \centering
    \includegraphics{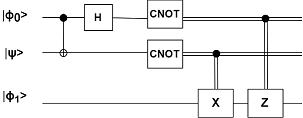}
    \caption{Informal description of Protocol}
    \label{Figure: 3.8 }
\end{figure}

On the other hand, the protocol has been described informally in plain English. The revolutions of the previous century have simplified certain algebras. Several non-simple algebras still wait to be simplified such as Algebra for modelling Quantum Teleportation protocol. In previous approaches, algebraic approach for teleportation was a little complex and demands a thorough understanding of the Algebra they use to model the teleportation protocol. Our research effort is to formalize the protocol in an easy to understand and simplified way. This would also helps us in verifying the correctness of quantum teleportation protocol using ePAlg syntax. The previous approaches mentioned in literature (e.g. \cite{29}) have tried to formulate teleportation as a program which can be used as a tool to implement a specification in this way $\psi^{'} =  \phi$. We have observed that the given specification can also be tested with a program which requires transferring the qubit through a quantum channel that needs not be the teleportation. Moreover, the specification has not explained the sending or receiving of the two classical bit that is the integral part in defining the teleportation process. For this communication, we also require a pair of maximally entangled qubit.
\\
As an illustration of the algebra, we give details of the teleportation protocol of \cite{2}. We write down the specification for Quantum Teleportation protocol and then write programs to satisfy these specifications. 

\textbf{\subsection{Specification}}
The formal specification of quantum teleportation has been given as:
\begin{equation*}
\begin{split}
    S \coloneqq var_{Alice} \, \wedge \,var_{Bob} \, \wedge \, \ket{\phi} \, \wedge \, \ket{\psi} \, \wedge \, \text{1 ebit} + \text{2 cbits} \geqslant  \text{1 qubit}
\end{split}
\end{equation*}
The specification  shows that the teleportation protocol works in a way that instead of using 1 qubit, we will be using 1 maximally entangled bit (1 ebit) and 2 classical bits (2 cbit) with a cost of sending 2 cbits from Alice to Bob for obtaining the Bob's z qubit. 
We have two processes $P^{Alice}$ and $Q^{Bob}$, and $\ket\phi$ = $\frac{\ket{00} + \ket{11}}{\sqrt{2}}$ is the entangled state that is partitioned among Alice and Bob as the first qubit $\ket{\phi_{0}}$ owns by Alice whereas the second qubit $\ket{\phi_{1}}$ belongs to Bob. Now Alice owns the two qubits, one from entangled state and other that needs to be sent to Bob $\ket\psi$ = $(\alpha\ket{0} + \beta\ket{1})$. Moreover, pq are the classical variables of type integer belongs to Alice that will store the classical outcomes obtained after applying measurement and will be sent on Channel and rs variables belongs to Bob which will receive the classical data from Alice and then apply unitary operations. And the, $var_{Alice} = \psi\phi_{0}, pq$ and $var_{Bob} = \phi_1,rs$ will be the set of variables that both of these processes own. 
\subsection{Algebraic Programs}
We now describe the programs $P^{Alice}$ and $Q^{Bob}$ that represents respectively, the sender as well as the receiver of teleportation protocol are represented by the program \textit{Teleport}.
\subsubsection{Program : $P^{Alice}$}
\begin{equation*}
\begin{split}
P^{Alice} \coloneqq \bigg[ x: Qubit,\, \, y: Qubit  \\
            x = \ket{\phi_{0}}, y = \ket{\psi} \\
               CNOT[var_{Alice}]  \\
              \Rightarrow CNOT[x,y] \\
              \Rightarrow CNOT[\ket{\phi_{0}}, \ket{\psi}]; \\
              H[x] \Rightarrow H[\ket{\phi_0} \\
                     \big[ p: Integer, q: Integer \\
                             measure[var_{Alice}, pq] \\
                               \Rightarrow measure[\{x,y\}, pq]\\
                               \Rightarrow measure[\ket{\phi_{0}\psi}, pq]; \\
             \big] \\
             & \bigg] 
\end{split}
\end{equation*}
\begin{figure}[t]
    \centering
    \includegraphics[width=8cm, height=8cm]{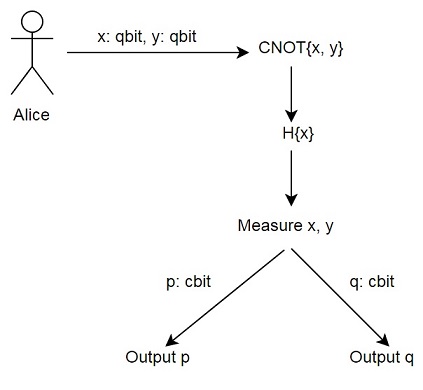}
    \caption{Behaviour of Alice}
    \label{Figure: 3.9 }
\end{figure}
\textbf{\subsubsection{Program : $Q^{Bob}$}}
\begin{equation*}
\begin{split}
    Q^{Bob} \coloneqq \Bigg[ z: Qubit\\
           \; \; \; \; \; \; z = \ket{\phi_{1}}\\
                \bigg[ r: Integer, \,\, s: Integer \\
           \; \; \; \; \; \; \; \ket{\phi_{1}} \Rightarrow Z[r].X[s].\ket{\phi_{1}} \\
             \bigg] \\
            &  \Bigg] \\
\end{split}
\end{equation*}
\begin{figure}[t]
    \centering
    \includegraphics[width=8cm, height=8cm]{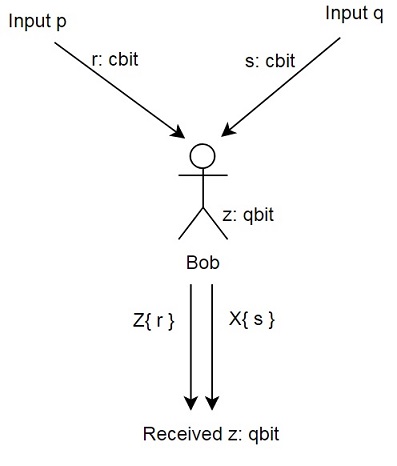}
    \caption{Behaviour of Bob}
    \label{Figure: 3.10}
\end{figure}
\textbf{\subsubsection{Program : BuildEPR}}
\begin{equation*}
\begin{split}
        BuildEPR \coloneqq \Bigg[  \ket{\Psi}: Qubit  \\
            \; \; \; \;   \bigg[ a: Qubit, \; b: Qubit \\
             \; \; \; \; \; \;     a = \ket{0},\; b = \ket{0} \\
                 H[a] \;\otimes\; I[b] \\
           \; \; \; \; \; \;        \Rightarrow H\ket{0} \otimes I\ket{0}; \\
                 CNOT[\;H[a]\;  \otimes  \; I[b]\;]  \\
                \Rightarrow CNOT[\;H\ket{0} \otimes I\ket{0}\;] \\
                \Rightarrow \ket{\Psi}
             \bigg] \\
             & \; \; \; \Bigg] \\
\end{split}
\end{equation*}
The above program creates EPR pair by applying Hadamard gate on first quantum state after applying CNOT gate with two $\ket{0}$ states as an input. The output state $\ket{\Psi}$ will be the EPR pair which is a classical example of an entangled pair.
\begin{figure}[t]
    \centering
    \includegraphics[width=6cm, height=13cm]{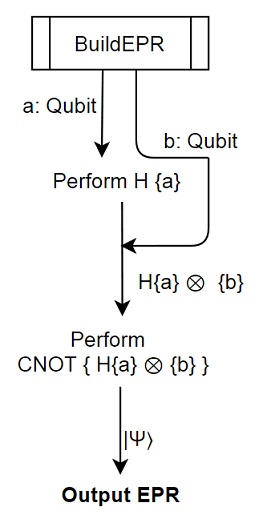}
    \caption{Behaviour of BuildEPR}
    \label{Figure:3.11 }
\end{figure}
\textbf{\subsubsection{Program : $QCOM$}}
\begin{equation*}
\begin{split}
P^{Alice}\;  \parallel_{\phi} \; Q^{Bob} \coloneqq 
             \;  Alice_{\phi_{0}\psi,pq}\; \parallel \;   Bob_{\phi_{1},rs} 
\end{split}
\end{equation*}
\newline
\textbf{\subsubsection{Program : Teleport}}
\begin{equation*}
\begin{split}
Teleport  \coloneqq  \Bigg[ \ket{\Psi}: Qubit  \\
              \bigg[ a: Qubit, \; b: Qubit \\
              \ket{\Psi} = BuildEPR[a,b] \\
           \ket{\Psi} = \ket{\phi_{0}} \; \otimes \; \ket{\phi_{1}} \\
           QCom \Rightarrow P^{Alice}  \; \parallel_{\phi} \, \, Q^{Bob} \\
             \bigg] \\
             & \Bigg] \\
\end{split}
\end{equation*}
The program \textbf{Teleport} will also be able to demonstrate that teleportation can get the z-qubit of Bob, if it has the real state of x-qubit of Alice, by sending only 2-bits from Alice towards Bob. \cite{3}.

\textbf{\section{Formally Modelling of Quantum Communication}}
Our main objective is to improve some algebraic representations for modeling the communication between two quantum processes. Before modeling the communication between quantum processes, we must accept that the quantum algorithms are made up of classical and quantum parts. They require cooperation between quantum and classical computation. Also, we know that quantum computation is probabilistic by nature and the results of these computations are checked and verified by the classical part. In case the classical result of the quantum computation is incorrect, we have to repeat this computation until we get the correct results. Quantum teleportation best exhibits the cooperation between classical and quantum computation. Through our upgraded process algebra, we intend to improve the previous version of communication transition rules between classical and quantum systems just to model this cooperation.
\\
\\
\textbf{\subsection{Improved Transition Rules for Modelling Communication}}
This section is subjected to extend some of the existing transition rules of communication between two quantum processes defined in \cite{3}\cite{34}. For communication, we have rewritten quantum and classical input, output transition rules here again with reasons on why previously written transition rules need improvement.

\textbf{\subsection{Classical Communication Transition Rules}}
In order to transfer classical data between two quantum processes, we have extended the syntax of classical value-passing \cite{2} and also modeled the transition rules for communication between two systems. 
\textbf{\subsubsection{C-IN}}
 $$ \mbox{\huge $ \frac{}{g?x.P \textbackslash C \quad \xrightarrow{g?v} \quad P \textbackslash C^{'}}$} $$ 
  where 
\begin{itemize}
 \item   $ C = \langle \; s, q = \rho, f \; \rangle $ 
 \item   $ C^{'} = \langle \; s, q = \rho, f \cup \{x \rightarrow{v} \}\; \rangle $ 
 \item   $ x \; \epsilon \;  Var(s) $ ,  $ x \; \epsilon \;$ Nat(Integer type)   and  $ v \epsilon N$ \;\text{N: Set of Natural numbers} 
\end{itemize}
\textbf{\subsubsection{C-OUT:}}
  $$ \mbox{\huge $ \frac{}{g!v.P \textbackslash C \quad \xrightarrow{g!v} \quad P \textbackslash C}$} $$ 
  where 
\begin{itemize}
 \item   $ C = \langle \; s, q = \rho, f \; \rangle $ 
 \item   $ v \epsilon N$ \;\text{N: Set of Natural numbers} 
\end{itemize} 
\textbf{\subsubsection{C-COM:} }
\begin{center}
\scalebox{1.4}{$$\mbox{$\frac{P\textbackslash C \, \xrightarrow{g?v} \quad P^{'} \textbackslash C \quad \quad Q\textbackslash C \, \xrightarrow{g!v} \quad Q^{'} \textbackslash C }{P \parallel Q \textbackslash C \, \xrightarrow{\tau} \quad P^{'} \parallel Q^{'}  \textbackslash C}$}$$} 
\scalebox{1.4}{$$\mbox{$\frac{P\textbackslash C \, \xrightarrow{g!v} \quad P^{'} \textbackslash C \quad \quad Q\textbackslash C \, \xrightarrow{g?v} \quad Q^{'} \textbackslash C }{P \parallel Q \textbackslash C \, \xrightarrow{\tau} \quad P^{'} \parallel Q^{'}  \textbackslash C}$}$$}
\end{center} 
where context of both systems remains unchanged.
These transition rules are similar to those defined in communicating data between two classical systems in classical process algebra{\cite{2}}. Contexts do not change in these transition rules as the information being modified is classical by nature and and also the quantum state of the accompanying systems remains unchanged. 
\textbf{\subsection{Quantum Communication Transition Rules}}
Communication between two quantum systems defines cooperation by classical information and quantum information. We have improved some important input and output transition rules here. Using these transition rules, we have written a rule that can define the communication between two quantum systems using parallel composition operator from process algebra \cite{2}.
\textbf{\subsubsection{Q-IN1:}}
  $$ \mbox{\huge $ \frac{}{g?x.P \textbackslash C \quad \xrightarrow{g?v: \sigma} \quad P \textbackslash C^{'}}$} $$  \\
where 
\begin{itemize}
 \item   $ C = \langle \; s, q = \rho, f \; \rangle $ 
 \item   $ C^{'} = \langle \; s, x.q = \rho \otimes \sigma , f \; \rangle $ 
 \item   $ \sigma \; \epsilon \; \mathcal{D(H}_2) $
 \item   $ x \; \epsilon \;  Var(s) $ ,  $ x \; \epsilon \; q $  and  $ v \notin q $
\end{itemize} 
\textbf{\subsubsection{Q-IN2:} }
  $$ \mbox{\huge $ \frac{}{g?x.P \textbackslash C \quad \xrightarrow{g?v} \quad P \textbackslash C}$} $$\\ 
where 
\begin{itemize}
 \item   $ C = \langle \; s, q = \rho, f \; \rangle $ 
 \item   $ x \; \epsilon \;  Var(s) $ ,  $ x \; \epsilon \; q $  and  $ v \epsilon q - \{ g?x.P \}$
\end{itemize} 
In previous work \cite{3}\cite{34}, the quantum input rule (Q-IN1) has been defined in a similar way, but that rule works only when the quantum system being input (here quantum variable v is representing input system) must not be in entangled state or is not correlated classically with the quantum system represented by x. As the entanglement is a physical phenomenon where the state of different quantum systems can be found entangled in such a way that the state of any of the two quantum systems cannot be measured independently of the state of other system even if they are physically light years away. The below mentioned rule clearly models the statement by this process algebra: \\   
$$ \mbox{\huge $ \frac{}{g?x.P \space \textbackslash C \quad \xrightarrow{g?v: \rho^{'}} \quad P \textbackslash C^{'}}$} $$ \\
where 
\begin{itemize}
 \item   $ C = \langle \; s, q = \rho, f \; \rangle $ 
 \item   $ C^{'} = \langle \; s, q = \sigma , f \; \rangle $ 
 \item   $ \sigma \; \epsilon \; \mathcal{D(H}_2) $
 \item   $ x \; \epsilon \;  Var(s) $ ,  $ x \; \epsilon \; q $  and  $ v \notin q $
 \item   $ C^{'} = \langle \; s, Tr_{q}(\sigma) = \rho^{'} , Tr_{v}(\sigma) = \rho ,f \; \rangle$ 
\end{itemize} 
We can model any kind of quantum data being input using this rule as no such change has been applied to the new quantum state other than this:  $Tr_{v}\sigma = \rho $ that says: the initial state of quantum systems will not be changed. The rule we derived above seems little absurd as it does not retain the property of being image finite \cite{2} from its previous states $ g?x.P\textbackslash C $ where context of this system is $ C = \langle x, q = \rho, f \rangle $ and the action it applies $ g?v:\rho^{'} $, there are never-ending numerous determined configurations fulfilling this rule.
\\
Taking in view the above discussion, we have introduced these two quantum input transition rules which explain the situations where the qubit is being inputted from outside the context of the system and within the context, separately. The point to be noted here again is that the context of the system remains unchanged after applying quantum input rule-2.

The input system has already been explained within the context of that specific process. And the above-mentioned rule doesn't modify the quantum state of this entire system (stored in x) as the process is only referencing the quantum system.
\textbf{\subsubsection{Q-OUT:}}
  $$ \mbox{\huge $ \frac{}{g!x.P  \textbackslash C \quad \xrightarrow{  g!x   } \quad P \textbackslash C}$} $$  \\
where 
\begin{itemize}
 \item   $ C = \langle \; s, q = \rho, f \; \rangle $ 
 \item   $ x \; \epsilon \; Var(s) $ and $ x \; \epsilon \; q $
\end{itemize} 
The output rule proposed in \cite{3}\cite{34} is as follows:

  $$ \mbox{\huge $ \frac{}{g!x.P \textbackslash C \quad \xrightarrow{g!x} \quad P \textbackslash C^{'}}$} $$ \\
where 
\begin{itemize}
 \item   $ C = \langle \; s, q = \rho, f \; \rangle $ 
 \item   $ C^{'} = \langle \; s \textbackslash\{x\}, q\textbackslash\{x\} = Tr_{q/\{x\}}(\rho) , f \; \rangle $ 
 \item   $ x \; \epsilon \; Var(s) $  and  $ x \; \epsilon \; q $
\end{itemize}
The instinct behind this old version of output rule is that once the quantum state of the system is sent out (output) to another system then this state (represented by variable name x) can be removed from the quantum sequence list as defined in this part ($q\textbackslash\{x\}$ and also from the stack of variables (\textbackslash\{x\}) of running process. Further, $Tr_{q\{x\}}(\rho)$ is used to obtain the quantum state by applying partial trace on $\rho$ over the qubit value in x. This transition rule also implies that the context of the quantum system being output is also changed.
\\
For justification of our improved transition rule, we assume that the system being sent out is correlated or entangled with the systems remaining in the context and once the state is removed from the context then this cannot be referenced in future. Now, after some time, if we input the same system again which has just been removed from the context then this can be problematic as we do not have this system in our context and hence we cannot measure the entangled state anymore. So, the Q-Output rule presented here can help us prevent this problem to happen since we do not wish remove the state being sent out from the context, so we keep that information. In this way, the context of the current system would be untouched.
\textbf{\subsubsection{Q-COM:}} 
 \begin{center}
     
 \scalebox{1.4}{$$ \mbox{$\frac{P\textbackslash C \,\xrightarrow{g?v} \quad P^{'} \textbackslash C \quad \quad Q\textbackslash C \, \xrightarrow{g!v} \quad Q^{'} \textbackslash C}{P \parallel Q \textbackslash C \, \xrightarrow{\tau} \quad P^{'} \parallel Q^{'}  \textbackslash C}$}$$}
  \\
\scalebox{1.4}{$$ \mbox{$ \frac{P\textbackslash C \, \xrightarrow{g!v} \quad P^{'} \textbackslash C \quad \quad Q\textbackslash C \, \xrightarrow{g?v} \quad Q^{'} \textbackslash C }{P \parallel Q \textbackslash C \, \xrightarrow{\tau} \quad P^{'} \parallel Q^{'}  \textbackslash C}$} $$}
 \end{center}
This is a simple rule representing the communication between two quantum systems outside of the context (using Q-IN2), but this is not completely quantum by nature as quantum state of the system is neither sent nor any measurements have to be applied so the context of these two systems remains unchanged.
\textbf{\subsection{Examples}}
Some examples are presented below to elaborate the transition rules formulated in last section. In the below mentioned theorems,  U will be the set of unitary transformation:
\begin{equation*}
    U = \{H, CNot\}
\end{equation*}
where H will be the Hadamard transformation and CNot will be the "Controlled-Not" operation. In the following theorems, we perform the Hadamard transformation by using our transition rules and explain how quantum systems are passed between two quantum processes even if they are entangled. After that, we see how the quantum systems work when we perform the unitary transformation. 
\begin{theorem}
If 
\begin{equation*}
P = g?y.P^{'} \quad \, , \, Q = g!x.Q^{'}    
\end{equation*}
and 
\begin{equation*}
R \quad = \quad (P\parallel Q) \textbackslash g  \; 
\text{where} \; y \notin q - (g!x.Q^{'})
\end{equation*}
Then 
\begin{equation*}
    T_{1}\, = \, R \textbackslash C \quad \xrightarrow{\tau} \quad (P^{'}\{x/y\} \parallel Q^{'}) \textbackslash g, C
\end{equation*}
\\
\\
\textbf{Proof:}
In the above transition $T_{1}$, the quantum system x is being sent out from the process Q to process P through quantum channel g. Note that, the context of the system remains unchanged as density operator $\rho$ only represents the quantum state of the output system x and doesn't store any other quantum information such as the position of the system. So, the above transition is valid and support the correctness of our communication rule.
\end{theorem}
\begin{theorem}
If 
\begin{equation*}
P_{1} = g?y.H[y].P_{1}^{'} \quad  , \quad  Q_{1} = (P_{1} \parallel Q) \textbackslash g  
\end{equation*}
and 
\begin{equation*}
\rho \,= \,\ket{1}_{x}\bra{1} \otimes \rho^{'}
\end{equation*}
\begin{equation*}
\text{where} \quad \quad y \notin q\; -\; (g!x.P_{1}^{'}), \; \;  \rho^{'}\, \in\, \mathcal{D(H}_{Var-\{x\}})
\end{equation*} 
Then 
\begin{equation*}
    T_{2} \, = \, Q_{1} \textbackslash C \quad \xrightarrow{\tau} \quad 
   (H[x].P_{1}^{'}\{x/y\} \parallel Q^{'}) \textbackslash g, C
\end{equation*}
\begin{equation*}
  \quad \quad   \quad \xrightarrow{H[x]} \quad (P_{1}^{'}\{x/y\} \parallel Q^{'}) \textbackslash g, C^{'}
\end{equation*}
Now the new state will be
\begin{equation*}
 \ket{-}_{x}\bra{-} \otimes \rho^{'}
\end{equation*}
\textbf{Proof:}
\\
\\
In this transition $T_{2}$, the context of the system is changed as the quantum system x is being sent out from $Q$ to $P_{1}$ and then the Hadamard transformation is applied on state  $\rho$ of the quantum system x and state $\rho$ of the quantum system x is changed from $\ket{1}$ to $\ket{-}$.
\end{theorem}
\begin{theorem}
If 
\begin{equation*}
P_{2} = g?y.CNOT[y,z].P_{2}^{'} \quad  , \quad  Q_{2} = (P_{2} \parallel Q) \textbackslash g  
\end{equation*}
and 
\begin{equation*}
\gamma\,= \,\ket{-}_{x}\bra{-} \otimes\ket{1}_{z}\bra{1} \otimes \gamma^{'}
\end{equation*}
\begin{equation*}
\text{where} \quad \quad y \notin q\; -\; (g!x.P_{2}^{'}), \; \;  \gamma^{'}\, \in\, \mathcal{D(H}_{Var-\{x,z\}})
\end{equation*} 
Then 
\begin{equation*}
    T_{3} \, = \, Q_{2} \textbackslash C \quad \xrightarrow{\tau} \quad 
   (CNOT[x,z].P_{2}^{'}\{x/y\} \parallel Q^{'}) \textbackslash g, C
\end{equation*}
\begin{equation*}
  \quad \quad  \quad \xrightarrow{CNOT[x,z]} \quad (P_{2}^{'}\{x/y\} \parallel Q^{'}) \textbackslash g, C^{'}
\end{equation*}
Now the new state will be
\begin{equation*}
 \ket{\alpha_{11}}_{x}\bra{\alpha_{11}} \otimes \gamma^{'}
\end{equation*}
\textbf{Proof:}
\\
\\
In this transition $T_{3}$, again the context of the system is changed as the system x is being sent out from Q to $P_{2}$ then unitary operation CNOT applied on quantum state $$\gamma$$ of the system x along with the system z and quantum state of the physical system\-xz will be entangled. As, we can see before applying CNOT, we have the quantum system x and quantum system z as separable states but once we applied CNOT operation we get the entanglement between these two quantum systems.
\end{theorem}



\textbf{\section{Discussion and Conclusion}}
Complete formal descriptions of quantum algorithms that uses the principles of quantum mechanics must uses both classical as well as quantum computational components and utilize these to make the communication and cooperation work. Furthermore, modeling of distributed and concurrent quantum systems, quantum to quantum communication and quantum communication protocols that physically transfers the qubits from one location to another must also be taken into account. 

In this paper, we have discussed the theory of formal modeling of concurrent and distributed quantum processes and provided with some improved transition rules for checking the correctness of communication between two quantum processes. We have discussed the limitation of some of the communication transition rules of the previous notation algebraic notation of quantum processes. 

Motivated by Milner's classical process algebras, which give an extended framework to formally model the cooperation between processes, Lalire have had introduced an algebraic notation for communicating processes which is named as Quantum Process Algebra (QPAlg). This Quantum Process Algebra gives a homogeneous style to formal specification and modeling of distributed and concurrent quantum systems consisting of both classical as well as quantum data. 
On the quantum part, QPAlg gives quantum variables, the operations on quantum variables such as applying quantum gates, pauli operators and the measurement observables, and also new types of communications including the quantum world. The operational semantics ensures that these quantum systems, their operations and the communication between these systems work as indicated by the postulates of quantum mechanics.

We have introduced an improved version of some algebraic representations for modelling the communication between two quantum processes and proved the formal model as a tool to check the correctness of practical concurrent quantum systems. Also, we have presented a direct approach for two quantum systems to communicate. We have identified a different approach to model the communication between distant quantum systems.We have further introduced the support of writing the specifications of working of distributed quantum systems. As an example, we formally specify the working of Quantum teleportation protocol and also provided the programs that best satisfies these specifications. Diagrams have also been added to best describe the flow and formal working of protocol. Moreover, we have introduced the concept of Entanglement  based instant messaging and mathematically perform the calculations. 

Further, in our examples we assumed to have two quantum systems and modeled two different examples of classical and quantum processes for checking the level of correctness of the communication between them. We have also applied our transition rules for some test examples where two processes are using the communication primitives of Quantum Process Algebra (QPAlg) provided that the communication is entirely based of the principles of quantum mechanics. We have elaborated the examples to describe the concurrent communication activities between these modeled systems.\\ 
On the theoretical side, there is a need to prove if the given program can prove the formal specification of teleportation protocol. We also aim to develop an approach for the development of a easy to understand quantum programming language and use our extended form of eQPAlg for modelling and specifying quantum superdense coding and BB84 protocol. On the practical side, there is a need to work on more practical examples for modeling cryptographic systems. Possibly a quantum cryptographic protocol can be modeled using the cryptographic protocol simulators.
\\
\\
\bibliography{researchpaper}

\end{document}